# IS SZILARD ENGINE REALLY BROKEN?

Srnivasa Rao. P

*Department of Mechanical Engineering, Vardhaman College of Engineering, Hyderabad, India*

## ABSTRACT

In the Second Law of Thermodynamics which is a macroscopic physical law that states that in any isolated system, entropy can only increase and order, or organization, can only decrease. This means that in the long run, energy will always dissipate and systems will always become less organized. James Maxwell proposed the idea of Maxwell's demon through his renowned thought experiment. Leo Szilard, on the other hand, demonstrated this idea with another thought experiment using a theoretical model of an information engine integrating Maxwell's demon and a simplified one-particle engine. This demonstration satisfactorily resolved the contradiction of the second law of thermodynamics. Landauer established the relation between the theory of information and thermodynamics, and proved that information is physical. However, recent publication contradicts the operation of such engine based on the principle of considering a piston without internal structure. In the present work, it is studied with the more fundamental idea of Vanderwalls' surface energy principle to substantiate the workability of such engine and has been observed that the engine can work with a similar established piston under defined conditions.

## INTRODUCTION

James Clerk Maxwell proposed a thought experiment in 1867 which has challenged the validation of the concept of Entropy and applicability of second law of thermodynamics [1][3][4]. Nearly 100 years ago Szilard conducted a thought experiment in the similar lines with 'a particle in a box' was the foundation stone in saving the second law of thermodynamics [5]. This thought experiment is the one which connects the concept of thermodynamic entropy with the concept of information [1][2]. Landaur likewise clearly proposes that information is inextricably linked to physical or tangible representation [2][4]. He also proposes that its strikingly more important to understand the information erasure than information recording. Moreover, Bennet and Penrose also proposed connectivity of information theory and Maxwell's Demon [6][7].

According to C. Maxwell by detecting and allowing the gas molecules to cross the trapdoor which takes no work to operate which partitions the gas chamber into two and separating the gas particles on the basis of its energy, this trapdoor is operated by an intelligent demon known as Maxwell's demon [8][9]. It essentially violates the second law of thermodynamics while doing so. While doing this the demon has to do something with information, the information about the energy and position of the discrete gas molecules of the chamber [10]. What the Maxwell's demon is doing is that, it records or gathers the information about each particle's position and other such details based on a threshold, the whole process decreases the entropy of the gas in the chamber [11][12]. The assumption is that the operation of the trapdoor is completely frictionless and consumes no energy. [13][14]

**SZILARD ENGINE**

In the year 1929, Leo Szilard proposed another thought experiment in the similar lines. According to Szilard, considering a situation of a single particle in a chamber (a box) similar to that of the Maxwell's case, with a separating wall which is movable like a piston. The consideration is represented as shown in the Fig.1 [1] [2] [3] [4].

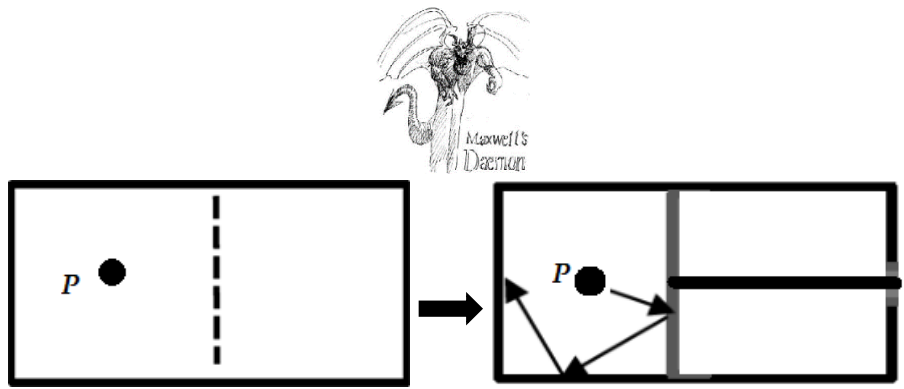

Figure.1. Szilard Engine: Demon finding particle 'P' in certain half of the chamber and replacing the partition with a movable piston which can do an external work

This setup sounds analogous to something like a bit of information, like finding a particle in a box. The work done by the piston due to the expansion of the particle in that chamber or enclosure can be derived as follows.

$$PV = nRT = \frac{N}{N_A}RT$$

$$P = \frac{R}{N_A}\frac{NT}{V} = k_B \frac{NT}{V}$$

As it is single molecule system, N=1 and Force = Pressure x Area

$$W = \int_{\frac{1}{2}}^{1} P dx = \int_{\frac{1}{2}}^{1} k_B \frac{T}{X} dX = k_B T \int_{\frac{1}{2}}^{1} \frac{1}{X} dX = k_B T ln(2)$$

Therefore, Szilard proposes that the system does the quantity of work which is equal to $k_B T \ln 2$ and hence the system returns to its original state. The Szilard engine so working from initial to final state interacting with single heat reservoir in a cyclic process, actually contradicts the second law of thermodynamics. However, this has been later resolved through Bennett, using Landauer principle [1][4][5][6]. He resolved this inconsistency, with an information erasure process as the final process in the Thermodynamic cycle comprising Szilard process. According to Landauer principle every bit of erasure takes at least $-k_B T ln(2)$, which has been experimentally verified using Brownian particles with a technique of optical trapping. It can be inferred from these statements as total work [8].

$$W_{total} = W_S + W_L = -k_B T ln(2) + k_B T ln(2) = 0$$

This conclusion satisfies the argument that an engine cannot function from a single heat source in accordance with the second law of Thermodynamics. [11][12]

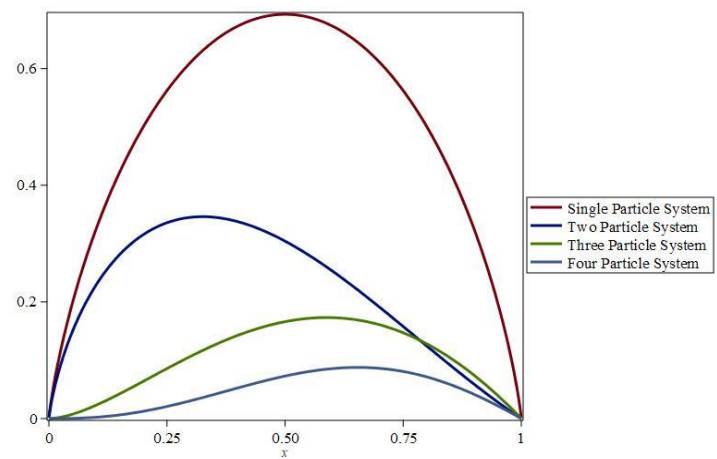

Fig. 2. Average Energy exchange by Szilard Engine with the partition

Fig.2 shows the average energy exchange that can be defined by the system. The system with single particle, two particle, three particle and four particles has been considered and energy exchange is always dropping or getting to an equilibrium due to the continuum of molecules established a defined energy level in the gas.

There has been an agreement on the functioning of Szilard engine and the information theory and later in the concept of information erasure [14]. Meanwhile, Quanmin Guo proposed that as the crucial step in the action of the Szilard engine is the isothermal expansion of a single particle system which cannot take place [15]. It was argued that the internal energy of the piston is not considered and such engine cannot do an external work as it is in the thermal equilibrium with the surroundings [18]. The thermal energy of the piston also has to be considered according to the proposed idea. The consideration of the fact of particle hitting the piston and moving it to do external work is not possible, because the exchange of momentum between the piston surface molecules and the free molecule is like a collision of particles as they are on the same energy levels. Hence it is not necessary to see for the compensation in the entropy as the engine itself cannot function [16][17].

In the case of a single molecule colliding with a piston, the highest amount of energy that can be transmitted from the molecule to the piston in a single collision is of the order of kT. Though it is a tiny amount of power, it may be reasoned that this energy is important for completing work by the piston, no matter how minute it is. At the core of Szilard engine, it is the most fundamental idea [16] [24]. However, it's important to recall that the energy within the piston varies by $\sqrt{3NkT}$, which is far more than the kT for a realistic piston. The energy variation within the piston is $10^{12}$ orders of magnitude bigger than kT for N to be on the order of $10^{23}$. Regardless of molecule's position in the box the energy of the piston fluctuates by an amount of $\sqrt{3NkT}$. As piston is assumed to be at equilibrium temperature with the experimental setup and therefore at T, the energy of the piston is unaffected by the addition or removal of energy of the system, this makes the system immune to the energy fluctuations. The average energy of each degree of freedom, according to the equipartition theorem, is $k_B T/2$. However, this equation only holds true on average for a large number of particles in equilibrium; each particle may have a different kinetic energy. The Boltzmann constant connects the microscopic and macroscopic worlds by linking the average energies of microscopic particles to the energies required to change the temperature in the macroscopic system, which is the most important element of the equation.

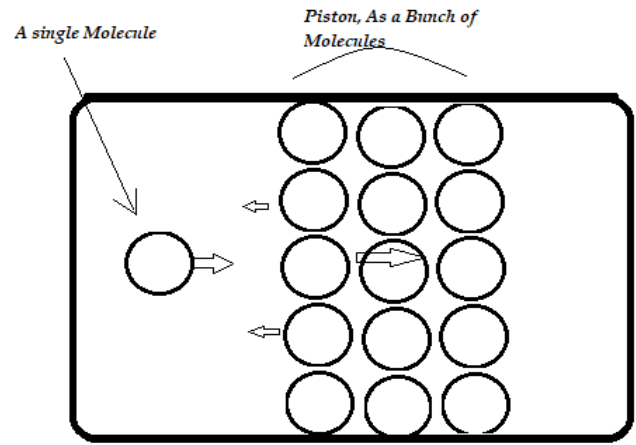

Figure.3. Szilard Engine with a piston (represented as bunch of molecules) and a single molecule

The fact of the energy fluctuations of individual atoms of the piston surface would also has a significant consequence viz. considering the arbitrary collision of free molecule and the group of molecules of the piston. It can be identified as kT is the amount of energy that could be transferred to the piston or from the piston to the molecule, however it can be inferred Statistically that after a large number of collisions the net energy transfer should be zero. The only way to transfer energy continuously from the enclosed chamber to the piston via single molecular expansion is to treat the piston as a mass with no internal structure. With an assumption of lighter piston, to the extent as nano-piston by arranging the molecules, the piston energy could be minimized approximately to $kT$ [18] [20].

It was concluded based on this fact that in the case of nano-piston, also the piston cannot be made functional as it could also be randomly moving in the confined box. This leads to a conclusion that such a model will not be sufficient to push the piston to the end of the box. The situation is analogous to two separate molecules in a box travelling at random. If the piston is assumed to be consisting of about $10^4$ molecules, it considerably takes almost 100 molecules to push it to one of the ends. But the practical problem is placing the piston, keeping the 100 molecules on one side. The probability of achieving is almost negligible. Consequently, leads to an un-resolvable predicament, leading to the failure of realizing such engine.

Thermodynamics and information are combined in a single framework. When a system begins and ends in a non-equilibrium state, the generalised second law is proven. The behaviour of the Szilard engine, which appears to defy the second law, has been explained. We've proven that, despite the fact that the system's energy remains unchanged, the non-equilibrium free energy appears to rise as measurements are taken. Hence one can extract energy in cyclic process from single heat bath using acquired information by measurement. However, erasure of stored information requires work to be done on system. A memory device can be used as an information reservoir which can be used to increase the performance of a device. We have formulated a simple model of autonomous information engine. We have found the system can act as an engine, refrigerator or an eraser. Even combination of any two is possible in some parameter space. We have achieved the efficiency of the engine to be greater than Carnot limit. The coefficient of performance of refrigerator also goes beyond the Carnot limit. Our findings are consistent with the generalized second law of thermodynamics along with information.

It is well known fact that forces of interaction between electrically neutral atoms, located at a distance '$D$' from one another which is large compared to their internal dimensions is inversely proportional to '$D^7$'. In the second approximation of perturbation theory applied to the electromagnetic force interactions there will be Van Der Waals forces getting generated, though valid for very small separations compared to the wavelengths.

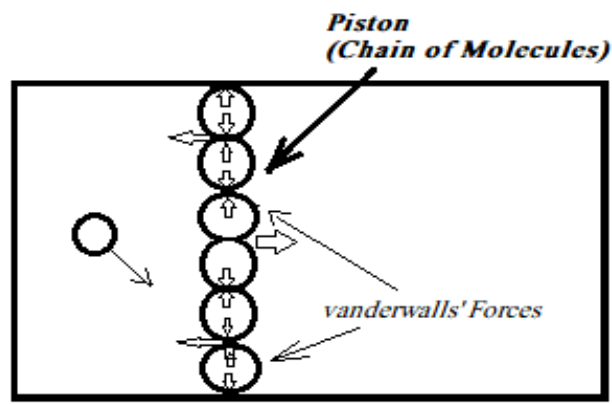

Figure.4. Schematic showing the interaction of molecule with the atoms of the piston. The velocities of molecule/atoms are the collision between the single molecule and an atom can result in energy transfer.

However, the direct measurement of molecular forces is not so easy to estimate in this kind of setup but apparently the works of Abrikosova et.al estimated such forces between the quartz plates separated roughly 0.1 to 0.4 μm. mentioning the possible optical limitations and force estimate approximations [21][22].

As the piston is placed two new surfaces have been created, in such a scenario it is necessary to justify the equilibrium, hence it is balanced by applying forces that are perpendicular to the previous surface, but the sums of these forces on the atoms on each side of the cut are zero and so do not affect the definition of the surface stress, as depicted in Figure.4[24][25][26].
The surface stress could be balanced either by external forces, e.g., the forces that must be applied to the edges of a plane or by a volume stress in the material below the surface, e.g. the hydrostatic pressure inside piston cylinder setup altogether [27].

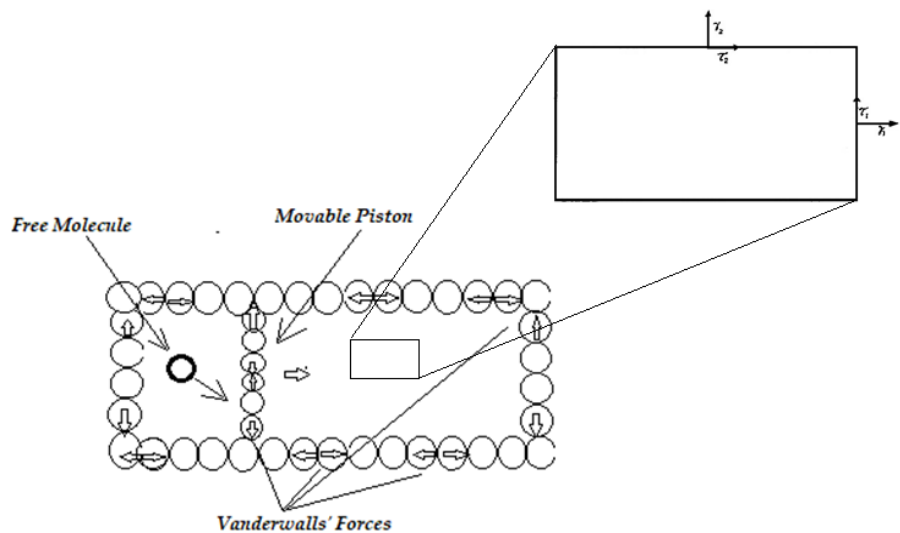

Figure.5. Schematic showing the interaction of molecule with the atoms of the piston. The velocities of molecule/atoms are the collision between the single molecule and an atom can result in energy transfer.

When external forces balance the surface stress, The molecule within a few atomic lengths of the point of application of the force gets rearranged, but the surface stress is considered a macroscopic term, therefore this local rearrangement has no comportment on the definition. The occurrence of large irreproducible forces of attraction is present due to the Van der Waals' forces as shown in the figure 5. The generation of shear stress and tangential stress on the thin piston cylinder arrangement can itself substantiate the theory of molecular attractive forces between solids.  according to the studies of

Derjaguin and others, there exists a significant disparity in published work in the area of Van der Waals force among macroscopic entities [21].

CONCLUSION

In this paper we have touched upon one of the most important phases in the development of information thermodynamics and evolution of information engine. The investigations and the findings in the gradual development of the concept of the role of thermodynamics in the information. But surprisingly the later work of Guo proposed that the working of such Szilard single molecule engine does not work. The piston will be unable to displace by the pressure from a single molecule. The critical point of confusion in the original concept of the Szilard engine is treating the piston as a mass without internal atomic structure as well as snubbing the fluctuation of its internal energy. We proposed that the Szilard engine with single molecule is workable as proposed in its conceptual version as the piston and the wall of the cylinder could have the force of bonding due to Van der waals which enables the sliding of the piston due to the collision of the single molecule which is one side of the piston. One of the bases for the consideration of electromagnetic fluctuations could also be seen in macroscopic applications that have been created and tested on the interaction of bodies brought close together. The effect of temperature on the behaviour of attractive forces in bodies would also affect the bodies come close and have such force of coalescence.